\documentclass[pdflatex,sn-mathphys-num]{sn-jnl}

\usepackage{graphicx}%
\usepackage{multirow}%
\usepackage{amsmath,amssymb,amsfonts}%
\usepackage{amsthm}%
\usepackage{mathrsfs}%
\usepackage[title]{appendix}%
\usepackage{xcolor}%
\usepackage{textcomp}%
\usepackage{manyfoot}%
\usepackage{booktabs}%
\usepackage{algorithm}%
\usepackage{algorithmicx}%
\usepackage{algpseudocode}%
\usepackage{listings}%
\usepackage{siunitx}

\theoremstyle{thmstyleone}%
\theoremstyle{thmstyletwo}%

\theoremstyle{thmstylethree}%

\begin{document}

\title[Article Title]{Fair Representation in Parliamentary Summaries: Measuring and Mitigating Inclusion Bias}


\author*[1,3]{\fnm{Eoghan} \sur{Cunningham}}\email{eoghan.cunningham@ucd.ie}

\author[2,3]{\fnm{James} \sur{Cross}}

\author[1,3]{\fnm{Derek} \sur{Greene}}

\affil*[1]{\orgdiv{School of Computer Science}, \orgname{University College Dublin}, \orgaddress{\city{Dublin}, \country{Ireland}}}

\affil[2]{\orgdiv{School of Politics and International Relations}, \orgname{University College Dublin}, \orgaddress{\city{Dublin} \country{Ireland}}}

\affil[3]{\orgdiv{Insight Centre for Data Analytics}, \orgname{University College Dublin}, \orgaddress{ \city{Dublin},  \country{Ireland}}}

\affil[]{\url{https://scholar.google.com/citations?user=7uqa6LkAAAAJ&hl=en}}


\abstract{The use of Large language models (LLMs) to summarise parliamentary proceedings presents a promising means of increasing the accessibility of democratic participation. However, as these systems increasingly mediate access to political information -- filtering and framing content before it reaches users -- there are important fairness considerations to address. In this work, we evaluate 5 LLMs (both proprietary and open-weight) in the summarisation of plenary debates from the European Parliament to investigate the representational biases that emerge in this context.
We develop an attribution-aware evaluation framework to measure speaker-level inclusion and mis-representation in debate summaries. 
Across all models and experiments, we find that speakers are less accurately represented in the final summary on the basis of (i) their speaking-order (speeches in the middle of the debate were systematically excluded), (ii) language spoken (non-English speakers were less faithfully represented), and (iii) political affiliations (better outcomes for left-of-centre parties). We further show how biases in these contexts can be decomposed to distinguish inclusion bias (systematic omission) from hallucination bias (systematic misrepresentation), and explore the effect of different mitigation strategies. Prompting strategies do not affect these biases. Instead, we propose a hierarchical summarisation method that decomposes the task into simpler extraction and aggregation steps, which we show significantly improves the positional/speaking-order bias across all models. 
These findings underscore the need for domain-sensitive evaluation metrics and ethical oversight in the deployment of LLMs for multilingual democratic applications.}




\maketitle

\section{Introduction}

Public understanding of parliamentary activities is worryingly low, despite their fundamental role in representative democracy. Research consistently demonstrates that citizens struggle to comprehend parliamentary activities, and often view parliaments as talking shops, where politicians grandstand, but do little else \cite{dalton2004democratic,hay2007we}. This points to a democratic deficit where public engagement with parliamentary processes is severely limited.
Citizens’ actions alone do not fully explain this \cite{lupia2016uninformed,mair2013ruling}. Information on parliamentary activities is generally made available to the public through plenary transcripts, legislative records, and detailed minutes of parliamentary meetings. However, while this data is theoretically open, in practice it is often not presented in a user-friendly or easily accessible format. As a result, citizens rarely take the time to engage with it.

The use of large language models (LLMs) for automated summarisation of parliamentary debates offers a promising way to address the accessibility gap and enhance citizens' engagement with democratic institutions \cite{combaz2025ai}. However, for such approaches to be effective, it is essential that the generated summaries accurately reflect both the content and the speakers involved. This raises fundamental questions of fairness: whose voices are included in summaries, whose positions are faithfully represented, and who is omitted entirely? This paper investigates these challenges using plenary speeches from the European Parliament.

LLMs can provide an effective means of summarising text \cite{zhang25survey}. However, alongside challenges such as summary fidelity \cite{fabbri2021summeval} and knowledge hallucination \cite{roller2020recipes}, the tendency of LLMs to omit details or overgeneralise in summaries has raised concerns \cite{peters2025generalization}, and LLMs have been shown to exhibit biases that pose fairness challenges in the specific context of summarising parliamentary proceedings. First, language models demonstrate uneven utilisation of information across their input context, known as the ``\textit{lost-in-the-middle}'' problem, or the ``\textit{middle curse}'' \cite{liu2024lost,ravaut2024context}. For parliamentary debates, this positional bias means speakers may be systematically under-represented according to \textit{when} they speak, irrespective of the substance of their contributions. Second, there is growing evidence of \textit{social biases} where language models demonstrate disparate treatment or outcomes between social groups \cite{bartl2024showgirls,gallegos2024biasa,mei2023bias}. In the context of summarising political texts, such biases may manifest along partisan lines \cite{huang2024biasa}. While there is growing evidence that LLMs demonstrate political bias during question answering tasks \cite{rettenberger2025assessing, rozado2024political, noels2025what}, similar biases remain under-explored in summarisation tasks; where speakers from different political groups may experience unequal outcomes in inclusion/omission or misrepresentation/hallucination \cite{steen2024bias}. Third, while contemporary LLMs have been shown to be effective in cross-lingual summarisation tasks \cite{park2024lowresource}, content in low-resource languages may receive less faithful representation than content in better resourced languages \cite{guo2024teaching,hangya2022improving}. This dimension of bias has received limited attention in the summarisation literature, yet is directly relevant to multilingual settings such as the European Parliament, which contains 24 official languages.

Standard summarisation evaluation metrics do not adequately capture these fairness requirements. For instance, a summary that reverses a speaker's position from opposition to support can still receive high scores if the semantic content is otherwise preserved. To address this, we develop an attribution-aware evaluation framework using reconstruction functions that extract speaker-specific content from summaries, enabling measurement of both inclusion and misattribution at the speaker level. We further investigate whether hierarchical summarisation approaches -- central to multi-document summarisation for decades -- can mitigate these biases when adapted to the LLM context with domain-informed structure.


The contributions of this paper are as follows:
\begin{enumerate}
\item An attribution-aware evaluation framework that distinguishes inclusion bias from hallucination bias, extending Steen and Markert's framework to multi-speaker parliamentary debates.
\item Evidence that hierarchical summarisation methods mitigate positional bias relative to zero-shot approaches.
\item Novel analysis of partisan bias in parliamentary summarisation, showing effects primarily related to inclusion (i.e., recall gaps) rather than misrepresentation.
\item Systematic examination of cross-lingual bias by language resource level in summarisation.
\end{enumerate}
For AI-generated summaries to serve democratic accountability, they must represent parliamentary voices equitably. In other words, not systematically favour speakers based on when they speak, which party they belong to, or which language they use. Our findings highlight both the promise and the limitations of current approaches: hierarchical summarisation methods substantially reduce positional bias, but partisan and cross-lingual disparities persist. More broadly, this work offers methodologies for evaluating fairness in multi-document summarisation systems where balanced source representation and accurate attribution are essential.

\section{Background}

\subsection{Multi-Document Summarisation}
\label{sec:background_mds}

Text summarisation is the process of automatically distilling the most important information from a source text to produce an abridged version for a particular user or task \cite{zhang25survey,mani1999advances}.
While early approaches to automatic summarisation were predominantly \textit{extractive}, relying on directly copying segments from the source text \cite{carbonell1998mmr}, many use cases benefit from \textit{abstractive} summaries that rephrase the original content. Notably, recent advances in LLMs have enabled the generation of more natural and coherent summaries that better resemble human writing, making them well-suited for such tasks \cite{zhang25survey}. Multi-Document Summarisation (MDS) presents additional challenges beyond single document summarisation, requiring systems to identify salient information across multiple sources while managing redundancy and potential contradictions. 

\subsubsection{Hierarchical Summarisation}
Hierarchical Summarisation, where content is first processed at a local level before global aggregation, is a popular method for tackling the information complexity of MDS \cite{radev2004mead,liu2019hierarchical,zhang2019hibert,chang2024booookscore}.
Hierarchical summarisation is analogous to the architecture of a convolutional neural network, in which early layers extract lower level features that are subsequently aggregated and contrasted in later layers to produce higher level representations \cite{zheng2024harnessing}. 
Early approaches relied on cluster-based methods that identified centroid concepts and built summaries around them \cite{radev2004mead}. With the advent of neural methods, hierarchical architectures became increasingly sophisticated and presented a means of summarising more information than could fit in the context windows of early language models. Liu \& Lapata introduced hierarchical transformers that encode documents separately before merging representations, demonstrating that such approaches better capture both local coherence and global dependencies than flat concatenation \cite{liu2019hierarchical}. Zhang et al. (HIBERT) showed that hierarchical document encoding with appropriate pre-training significantly improves extractive summarization \cite{zhang2019hibert}.

In recent years, the development of LLMs with extended context windows appeared to limit the technical necessity of hierarchical architectures to very long context summarisation tasks, such as when working with long-form books \cite{chang2024booookscore}. However, recent works have identified a critical limitation: models exhibit strong positional biases in how they utilise information in context \cite{liu2024lost, ravaut2024context, sun2025posumbench}.
In particular, many models ignore material from the middle of the source when generating summaries, a phenomenon sometimes referred to as the ``middle curse'' \cite{ravaut2024context,sun2025posumbench}.
Ravaut et al. investigated whether hierarchical approaches could mitigate this positional bias across a variety of summarisation tasks with varying levels of success \cite{ravaut2024context}. They found that hierarchical summarisation was effective for scientific publications -- which have a prescribed structure -- but harmful in all other domains. These findings suggest that hierarchical approaches may be effective for summarisation when the hierarchy reflects meaningful semantic or functional boundaries rather than arbitrary textual divisions.
In our experiments on parliamentary debates, we investigate whether imposing a domain-informed structure will allow hierarchical methods to mitigate positional biases in debate summaries. 

\subsection{Evaluating Summaries}
\label{sec:background_eval}

Evaluation methods for text summarisation have been developed to automatically assess how effectively a summary captures the important content of a longer text while remaining concise, coherent, and faithful to the source material. 
Traditional approaches have focused on explicit overlaps between the summary and the source material. \textit{ROUGE} \cite{lin2004rouge} measures n-gram \textit{recall} -- the proportion of n-grams in the summary that were present in the source material. 
Similarly, \textit{BLEU} \cite{papineni2002bleu} measures the \textit{precision} of the n-grams in the summary compared to the source. As these methods rely on n-grams comparisons, they require that the summary text contains exact matches with the source text. As a result, they penalise paraphrased or \textit{abstractive} summaries, such as those generated by modern LLM-driven summarisation methods. Moreover, such measures have previously been shown to correlate poorly with human judgments of summaries \cite{nenkova2006summarization}. 

%
To evaluate abstractive summaries, embedding-based similarity measures have been developed to address the semantic limitations of the earlier methods that relied on n-gram overlap. For example, \textit{BERTScore} \cite{zhang2020bert} measures \textit{precision} and \textit{recall} separately, by comparing the semantic similarity of tokens in the summary and source material. Specifically, a high \textit{BERTScore} precision ($P_{\text{BERT}}$) indicates that each token/term in the summary is a strong semantic match for a token in the source (indicating a low level of added information or hallucination). Conversely, a high recall ($R_{\text{BERT}}$) indicates that tokens/terms in the source are all matched by tokens in the summary -- indicating that the summary covers the content in the original document. 
%
More recent work has explored \textit{factual consistency} as an evaluation criterion, which considers whether a summary accurately conveys the meaning of the source text without introducing contradictions \cite{wang2020fact}. Such approaches have been applied in the context of question answering (QA) \cite{scialom2021questeval} and natural language inference (NLI) \cite{falke2019ranking}. However, their effectiveness relies heavily on the quality of the underlying QA and NLI models.


When summarising parliamentary debates, particular attention must be given to accurately attributing arguments, positions, and proposals to the correct speakers. Existing evaluation methods 
do not adequately penalise the misattribution of content to speakers. For instance, standard metrics can assign high scores to summaries that reverse a speakers position (from opposition to support) or ignore them entirely (provided another speaker with a similar position is present), because these methods measure an overall semantic or lexical overlap, rather than speaker-specific accuracy. Later in Section \ref{sec:method_evaluation}, we demonstrate this limitation with examples, and describe our framework for evaluating debate summaries where we address this evaluation gap. 

\subsection{Types of Bias in LLMs}
\label{sec:background_bias}

As discussed in Section~\ref{sec:background_mds}, there is existing research that finds that LLMs do not attend equally to all regions of the input context when generating outputs. 
For parliamentary debates, this means speakers may be under-represented based on \textit{when} they speak rather than \textit{what} they say. Beyond this algorithmic, positional bias, LLMs have been shown to exhibit \textit{cognitive} biases, which affect their performance in decision making tasks \cite{lou2025anchoring}, and \textit{social} biases \cite{bartl2024showgirls,mei2023bias}, which (more relevant to this study) can be defined as disparate treatment or outcomes between social groups \cite{gallegos2024biasa}.

In the context of \textit{political information}, studies have identified that many LLMs demonstrate partisanship in (i) how they answer questions regarding policy preferences \cite{rozado2024political,rettenberger2025assessing}, and (ii) what information they exclude or censor when providing information about political figures \cite{noels2025what}.
In the context of \textit{summarisation}, Steen and Markert \cite{steen2024bias} distinguish between different manifestations of bias: \textit{inclusion} bias, where certain groups are systematically omitted from summaries, and \textit{hallucination} bias, where content is fabricated or misattributed to members of certain groups. This distinction informs our experiments, and how we interpret results: inclusion bias reflects selective attention (whose voices are heard), while hallucination bias reflects misrepresentation (whose positions are distorted). While research on political institutions (like parliaments) is limited, an important study using smaller pre-trained models (GPT-2, BART, T5) found that they demonstrate an inclusion bias in favour of left-of-centre positions when summarising opinions from social media \cite{huang2024biasa}. 

A further dimension relevant to multilingual settings such as the European Parliament is cross-lingual bias. While LLMs have been shown to achieve state-of-the-art performance in cross-lingual summarisations \cite{park2024lowresource}, they are known to perform less reliably on low-resource languages \cite{hangya2022improving,guo2024teaching}. The implications for summarisation in multilingual settings, where speakers using different languages may receive unequal representation, remain underexplored.

Throughout our experiments, we evaluate each of these three dimensions of bias under the definitions of \textit{group fairness} proposed by Gallegos et al. \cite{gallegos2024biasa}, requiring that summaries represent speakers with equal accuracy regardless of party, language or speaking-order. 

\section{Methods}

\subsection{Generating Debate Summaries}
\label{sec:method_summarisation}

As previously noted, summarising parliamentary debates presents distinct challenges. Unlike many traditional summarisation tasks, it requires not only an accurate representation of the arguments made but also precise attribution to the correct speakers. Moreover, to promote trust, fairness, and transparency, the content of the summary must be clearly traceable to the original source material.

Formally, a debate is represented by an ordered sequence of $n$ interventions made by $m$ different speakers:
\begin{equation}
I = \{i_1^1, i_2^2, i_3^3, \ldots, i_m^n\}
\end{equation}
where $i_j^k$ is a contribution made by speaker $j$, that was the $k$-th contribution or intervention in the debate.

\begin{figure}
\includegraphics[width=0.95\linewidth]{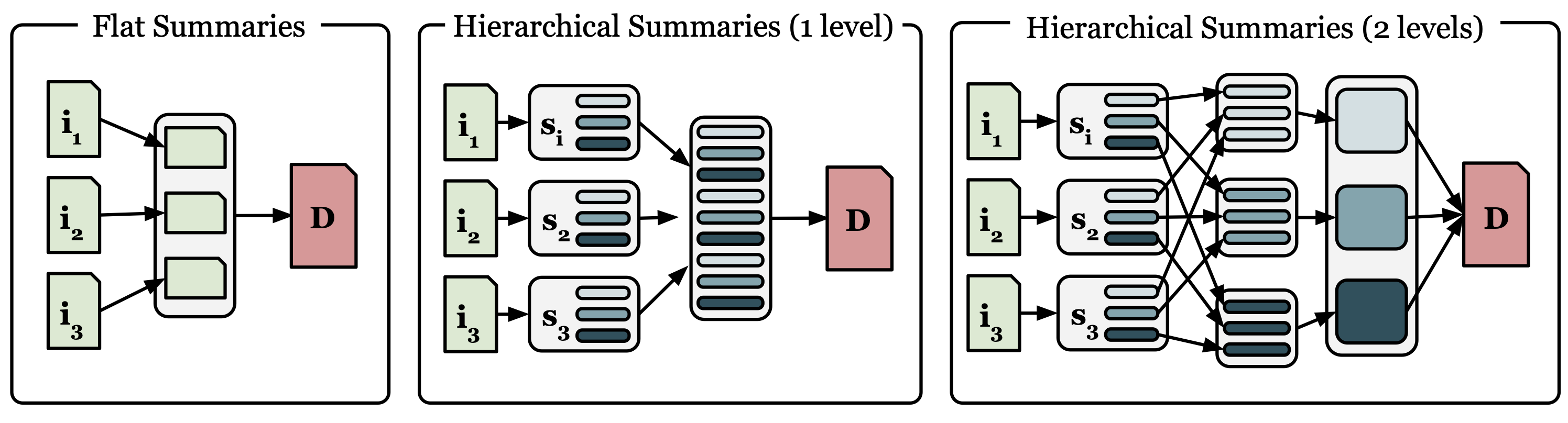}
\caption{{\textbf{Debate summarisation workflows.}} Flat debates summaries concatenate all interventions into a single document and summarise. Hierarchical summaries use a multi-step process where each intervention in the debate is summarised independently before aggregation. In the 2-level hierarchical approach, we extract information from intermediate intervention to aggregate thematically (i.e., ``summarise all arguments/issues/proposals'') prior to the final aggregation.} \label{fig:generating_summaries}
\end{figure}

We investigate how different levels of hierarchical structure affect summary quality and fairness. Specifically, we compare four approaches that we illustrate in \ref{fig:generating_summaries}:
\begin{enumerate}
\item \textbf{Flat summarisation}: Directly generating a debate summary from all interventions
\item \textbf{Flat summarisation + Prompt}: Directly generating a debate summary from all interventions, with explicit prompting to ``give equal attention to all speakers in your summary''
\item \textbf{One-level hierarchical summarisation}: Creating intervention summaries, then aggregating them into a debate summary
\item \textbf{Two-level hierarchical summarisation}: Creating intervention summaries, applying an additional intermediate aggregation step, then generating the final debate summary
\end{enumerate}

\subsubsection{Flat Debate Summaries}

In the flat approach, we directly generate a debate summary $D$ from the complete set of interventions:
\begin{equation}
D = f_{GEN}^{(0)}(I)
\end{equation}
The generator function $f_{GEN}^{(0)}$ takes the full debate text (all interventions $I$) as input and produces a debate summary in a single step. This serves as our baseline, representing the most straightforward application of LLMs to debate summarisation without any hierarchical structuring.
We evaluate two variants of the flat approach:
\begin{enumerate}
\item \textbf{Default}: Standard summarization prompt
\item \textbf{+ Prompt}: Includes an explicit instruction to ``give equal attention to all speakers in your summary''. Prompt engineering of this nature represents the simplest effort to mitigate attention biases and has been shown to be effective in certain contexts~\cite{ravaut2024context}.
\end{enumerate}
In our experiments, we implement this generator function using the LLMs listed in Section 4.2.

\subsubsection{One-Level Hierarchical Debate Summaries}

To implement hierarchical summarisation, we break the summarisation down into a series of extraction and aggregation steps.
First, create structured summaries of individual interventions. Each debate intervention is summarised by $f_{SUM}$, which maps an intervention to a structured summary:
\begin{equation}
s_i^j = f_{SUM}(i_i^j)
\end{equation}
Drawing on research from legislative bargaining and policy negotiation~\cite{thomson2006research,thomson2012new}, we decompose each intervention according to: the issues raised, positions taken, arguments presented, and proposals offered. While this literature typically focusses on measuring actors' policy positions and saliences for predictive modelling, we adapt these concepts into an explicit structural template for summarisation:

\begin{minipage}{0.98\textwidth}
\vskip\baselineskip
{\ttfamily\small
Headline: A concise, single-line summary of the speech.\\
Issue: A brief overview of any key issues raised by the speaker.\\
Position: Any stance or viewpoint expressed in the speech.\\
Argument: Any arguments used to justify the positions taken.\\
Proposal: Any proposals or policy actions mentioned by the speaker to address the issues raised.\\
Quotes: 2--3 representative quotes that capture the speaker's stance.\\
}
\end{minipage}

This structure provides the scaffolding for hierarchical summaries, constraining each step to a simpler, more tractable task. We rely on this existing structure from bargaining research to distinguish our approach from prior work applying hierarchical methods to LLM summarisation~\cite{chang2024booookscore}, where arbitrary chunking (e.g., by chapter) failed to mitigate positional bias. 
It is intended to focus the final summary on the substantive aspects of the debate interventions, specifically the issues raised, positions expressed, arguments presented, and proposals put forward by the speakers. Each component is semantically complete and comparable, providing simpler aggregation and comparison of relevant information. 

After the intervention summarisation step, a debate is represented as a set of structured summaries:
\begin{equation}
S = \{s_1^1, s_2^2, s_3^3, \ldots, s_m^n\}
\end{equation}
%
%
In the one-level hierarchical approach (illustrated in Figure~\ref{fig:generating_summaries}), the debate summary $D$ is produced by applying a generator function $f_{GEN}^{(1)}$ to the structured set of intervention summaries:
\begin{equation}
D = f_{GEN}^{(1)}(S)
\end{equation}



\subsubsection{Two-Level Hierarchical Debate Summaries}

The two-level hierarchical approach introduces an additional intermediate aggregation step. After creating intervention summaries, we apply a thematic aggregation function $f_{\text{THEME}}$ 
in which all positions, issues, arguments, and proposals are summarised independently and then aggregated into a final summary. Specifically, we let:
\begin{align}
T &= \{ \ f_{SUM}^{\text{theme}}(S) \ \forall \ \text{theme} \in \text{Themes} \ \} \\
D &= f_{GEN}^{(2)}(T)
\end{align}
where $T$ represents a set of intermediate themematic summaries, and $f_{GEN}^{(2)}$ produces the final debate summary. Using this structure, we perform the following steps:
\begin{enumerate}
\item Extract all positions from intervention summaries $S$ and create a summary of positions
\item Extract all issues from intervention summaries $S$ and create a summary of issues 
\item Extract all arguments from intervention summaries $S$ and create a summary of arguments
\item Extract all proposals from intervention summaries $S$ and create a summary of proposals 
\item Aggregate the thematic summaries from Steps 1--4 to produce the final debate summary $D$
\end{enumerate}

We demonstrate this process visually in Figure~\ref{fig:generating_summaries}. Hierarchical pipelines have been shown to be effective in summarising complex information, provided there is a clear organisational structure in the input~\cite{ravaut2024context,zhang2025summact,zheng2024harnessing}. By organising information along the substantive dimensions of legislative discourse (issue, position, argument, proposal) rather than arbitrary boundaries, we simplify the aggregation process by reducing the information to semantically complete, comparable chunks.




\subsection{Evaluating Debate Summaries}
\label{sec:method_evaluation}
The goal of summarisation is to generate a concise, coherent version of a longer text that preserves key information and enables quick understanding of the main points.
When working with parliamentary debates, we place an additional emphasis on the substantive aspects of debate and policy bargaining (i.e., issue, position, argument, proposals), and the reliable attribution of each of these concepts to the correct speakers. 
That is to say, when reading a debate summary, it should be clear \textit{what} was said, and \textit{who} said it. 
Many measures have been developed for the automatic evaluation of natural language summaries (see Section \ref{sec:background_eval}). 
Broadly speaking, when evaluating debate summaries these methods assess \textit{what} was said but they do not sufficiently penalise summaries for misattributing statements to different speakers. To demonstrate this gap, consider three speakers discussing office spending:

\begin{quote}
{\ttfamily
John: ``I strongly support the proposal to spend more money on doughnuts.''

Paul: ``I believe that the proposal is a waste of money.''

Ringo: ``I think the money should be spent on pastries.''
}
\end{quote}

In Table~\ref{tab:attribution_failure_example_1}, we show how \textit{the BERTScore} and \textit{ROUGE-2} evaluate simple summaries of this toy debate with different attribution failures. Both methods penalise paraphrasing more than speaker omission or misattribution.

\begin{table}[h]
\centering
\begin{tabular}{lcc}
\hline
Summary Type & \textit{BERTScore} & \textit{ROUGE-2} \\
\hline
Faithful summary & 0.79 & 0.86 \\
John/Paul opinion reversed & 0.72 & 0.58\\
Ringo completely omitted & 0.71 & 0.69 \\
Legitimate paraphrase & 0.62 & 0.07 \\
\hline
\end{tabular}
\caption{\textbf{\textit{BERTScore} and \textit{ROUGE-2} penalise paraphrasing more than speaker omission or attribution errors.}}
\label{tab:attribution_failure_example_1}
\end{table}

To focus our evaluations on the correct attribution of issues, positions, arguments, and proposals to speakers, and address this evaluation gap we must identify the regions of the debate summary that are relevant to each speaker. 
From each debate report $D$, we reconstruct the contributions of each speaker using a reconstructor function $f_{\text{REC}}$, where 
\[
\hat{s}_i^j = f_{\text{REC}}(D, \text{speaker}_j)
\]
represents a summary of the $i^{th}$ intervention in the debate (made by speaker $j$), that was \textit{extracted} (or \textit{reconstructed}) from information in the debate report $D$. Thus, we can evaluate the report (and the associated aggregator function) by comparing the reconstructed intervention summaries $\hat{S} = \{\hat{s}_1^1, \hat{s}_2^2, \dots, \hat{s}_n^m\}$, with the original interventions. Specifically, if a reconstructed summary $\hat{s}_i^j$ remains faithful to the initial intervention ${i_i}^j$, we conclude that the summarisation has attended to that speaker and their contribution accurately and, critically, that the contribution was clearly attributed to that speaker. Further, we can assess methodological and domain-specific biases in the generated report by comparing the fidelity of the interventions given different factors such as their order in the context (i.e., when they spoke during the debate), the language they spoke, or the political affiliations of the speaker (e.g.: party membership). 

\begin{figure}[t!]
\centering
\includegraphics[width=0.45\textwidth]{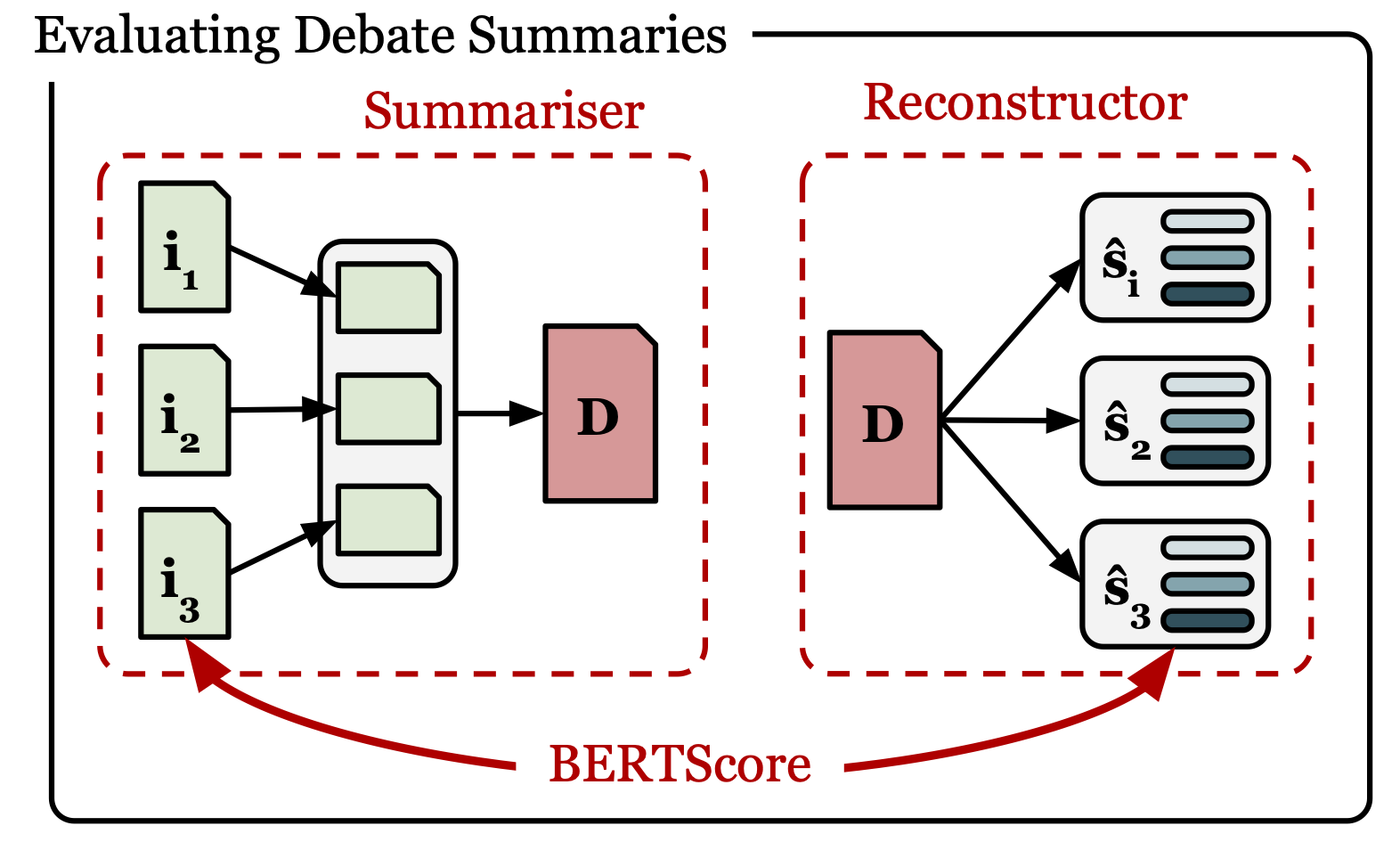}
\caption{{\textbf{Proposed evaluation workflow.}} The reconstructor function recreates a structured version of a speaker's intervention using the final debate summary as input. If the speaker's contribution can be accurately reconstructed, we conclude the summary attends to that speaker and their contributions accurately.}
\label{fig:fig2}
\end{figure}

We use \textit{BERTScore} \cite{zhang2020bert} to evaluate the accuracy (fidelity) of the reconstructed intervention summaries compared to the original speeches. As such, $\textit{BERTScore}(i_i^j, \hat{s_i}^j)$ measures how faithfully the substantive aspects of the intervention $i_i^j$ (i.e., issue, position, argument, proposals) were communicated in the debate summary and correctly attributed to the speaker. By measuring the intervention fidelity ($\frac{1}{n}\sum_{i}^{n}\textit{BERTScore}(i_i, \hat{s_i})$) for each intervention in a debate, as opposed to entire debate summary fidelity ($\textit{BERTScore}(I,D)$), we build attribution accuracy into our evaluation of fidelity. See Table~\ref{tab:attribution_failure_example_2} for this approach applied to the toy example from above. 
Further, as the \textit{BERTScore} metric is a combination of measures of \textit{BERTScore} recall and \textit{BERTScore} precision (see Section~\ref{sec:background_eval}), we use these metrics to operationalise Steen and Markert`s definitions of inclusion bias and hallucination bias respectively. 

\begin{table}[h]
\centering
\begin{tabular}{lcc}
\hline
Summary Type & $\textit{BERTScore}(I,D)$ & $\frac{1}{n}\sum_{i}^{n}\textit{BERTScore}(i_i, \hat{s_i})$ \\
\hline
Faithful summary & 0.79 & 0.77 \\
John/Paul positions reversed & 0.72 & 0.50 \\
Ringo completely omitted & 0.71 & 0.51 \\
Legitimate paraphrase & 0.62 & 0.56 \\
\hline
\end{tabular}
\caption{\textbf{Measuring \textit{BERTScore} speaker-by-speaker penalises speaker ommission and attribution errors.}}
\label{tab:attribution_failure_example_2}
\end{table}

The data and prompts used for all experiments are available with the code on {\ttfamily Github.com}\footnote{\url{https://github.com/parliview/debate-summ-cross-lingual}}.

\section{Experimental Setup}

\subsection{Data}

For our experiments we use a dataset of debates from the 6th European Parliament (EP)\footnote{\url{https://www.europarl.europa.eu/plenary/en/home.html}}, covering the period 2005 to 2006. 
We take a sample of 1,242 speeches (interventions) from 50 debates.
In order to evaluate the effects of bias in a many-to-one, cross-lingual summarisation setting, we require official translations of the debates alongside the original multi-lingual transcript.
Although the EP contains 26 official languages, and live interpretations are available to all attendees of plenary sessions, the European Parliament stopped making official translations of transcripts available after the 6th term. This is the reason we use debates sampled from the 2005--2006 period.
The final dataset comprises 2,484 interventions (the original speech and the official English translation) delivered by 708 speakers from 7 different party groups which generally consist of politicians with aligned ideologies. 

\subsection{Models}
\label{sec:experiments_models}
We use a range of LLMs to implement the summariser, generator and reconstructor functions in our experiments.

\subsubsection{Generating summaries.} For this process, we consider a selection of openly-available and proprietary models:
\begin{itemize}
    \item \textbf{Claude Sonnet 4} is a proprietary model from \textit{Anthropic}. In our experiments, we use {\ttfamily claude-4-sonnet-20250514}.
    \item \textbf{Claude Haiku 4.5} is a proprietary model from \textit{Anthropic}. In our experiments, we use {\ttfamily claude-haiku-4-5-20251001}.
    \item \textbf{Phi4} is a compact 14 billion parameter, open-weight model from \textit{Microsoft}. In our experiments, we use {\ttfamily Phi-4:14b} from \textit{Ollama}.
    \item \textbf{GPT-5} is a state-of-the-art proprietary model from \textit{OpenAI}. In our experiments, we use {\ttfamily gpt-5} via \textit{OpenAI} API.
    \item \textbf{Gemma3} is a lightweight 4 billion parameter, open-weight model from \textit{Google}. In our experiments, we use {\ttfamily gemma3:4b} from \textit{Ollama}.
\end{itemize}
For each model and summarisation approach (see Section~\ref{sec:method_summarisation}), we produce two summaries of every debate: one from the official English-language transcript and one from the original multilingual transcript. In total, this yields 2,000 debate summaries, which we use to examine three potential sources of inclusion bias: (i) when in the debate a speaker contributes, (ii) the language in which they speak, and (iii) their political affiliation.


\subsubsection{Evaluating summaries.} As the evaluations are the most costly aspect of our experiments, we make use of the open-weight model \textbf{Qwen3} from \textit{Alibaba}. Specifically, the function for extracting speakers' positions from  debate summaries ($f_{\text{REC}}$), is implemented using {\ttfamily qwen3:30b-a3b} provided by \textit{Ollama}, which employs mixture-of-experts to offer high performance with significantly fewer active parameters.
For clarity, and to avoid potential biases, we exclude this model from summary generation.


\subsection{Validation}

The reconstructor function ($f_{\text{REC}}$) is a critical stage in our evaluation pipeline. As such, we employ two measures to validate this aspect of the experiments. Firstly, we assess the effect of the choice of $f_{\text{REC}}$ model on our results. In other words, if we change the model we use to reconstruct/extract speaker interventions from debate summaries, would our results change significantly? To test this, we take a sample of 7 debate summaries generated using different generation methods. For each of the 147 different debate interventions, we reconstruct intervention summaries ($\hat{s}_i = f_{\text{REC}}(D, \text{speaker}_i)$) using four different models (\textit{Mistral-7b}, \textit{Phi-4}, \textit{Qwen3}, and \textit{Claude-Sonnet-3-7}). We then calculate the intervention fidelity ($\textit{BERTScore}(i_i, \hat{s_i})$) for all interventions, and measure the pairwise Pearson correlation between scores from all models. 
All model pairs showed high correlation in their scores, with the lowest at $r = 0.74$, indicating strong agreement across models. The correlation between \textit{Qwen3} and \textit{Claude-Sonnet-3-7} scores was the highest at $r = 0.86$. This suggests that the observed trends are largely robust to the choice of model used for reconstructing intervention summaries from debate summaries.

Next, we assess whether the chosen \textit{Qwen3} model can reliably extract all information relevant to a given speaker from the debate summaries. We select 30 speakers from a stratified sample of debates, models, and summarisation methods, and ask three expert human annotators to classify each extraction as \textit{accurate} and \textit{complete}. The codebook used for this task is available with the code on {\ttfamily Github.com}\footnote{\url{https://github.com/eoghancunn/position_extraction_annotation}}. Across all annotations, the accuracy of the extractions is 96\%. If we employ majority voting, the overall accuracy raises to 97.5\%. The average pairwise agreement between annotators is 93\%.


In conclusion, we find (i) that our reconstructor function $f_{REC}$ is reliably capable of extracting information about a speaker from a debate summary, (ii) that differences measured between the original intervention ($i_i$) and the reconstructed summary ($\hat{s}_i$) are artefacts of the summarisation method (and not the evaluations), and (iii) that our results are robust to our choice of model to implement $f_{REC}$.

\section{Results and Discussion}

We now evaluate three key factors which may influence inclusion bias in parliamentary debate summarisation. Specifically, we measure the accuracy with which speakers are represented in debate summaries according to:
\begin{enumerate}
    \item when they speak in the debate -- in Section~\ref{sec:results_order_bias}
    \item the language in which they speak -- in Section~\ref{sec:results_language_bias}
    \item their political affiliations -- in Section~\ref{sec:results_party_bias}
\end{enumerate}

\subsection{Lost in the Middle}
\label{sec:results_order_bias}

\begin{figure}[!t]
\centering
\includegraphics[width=\textwidth]{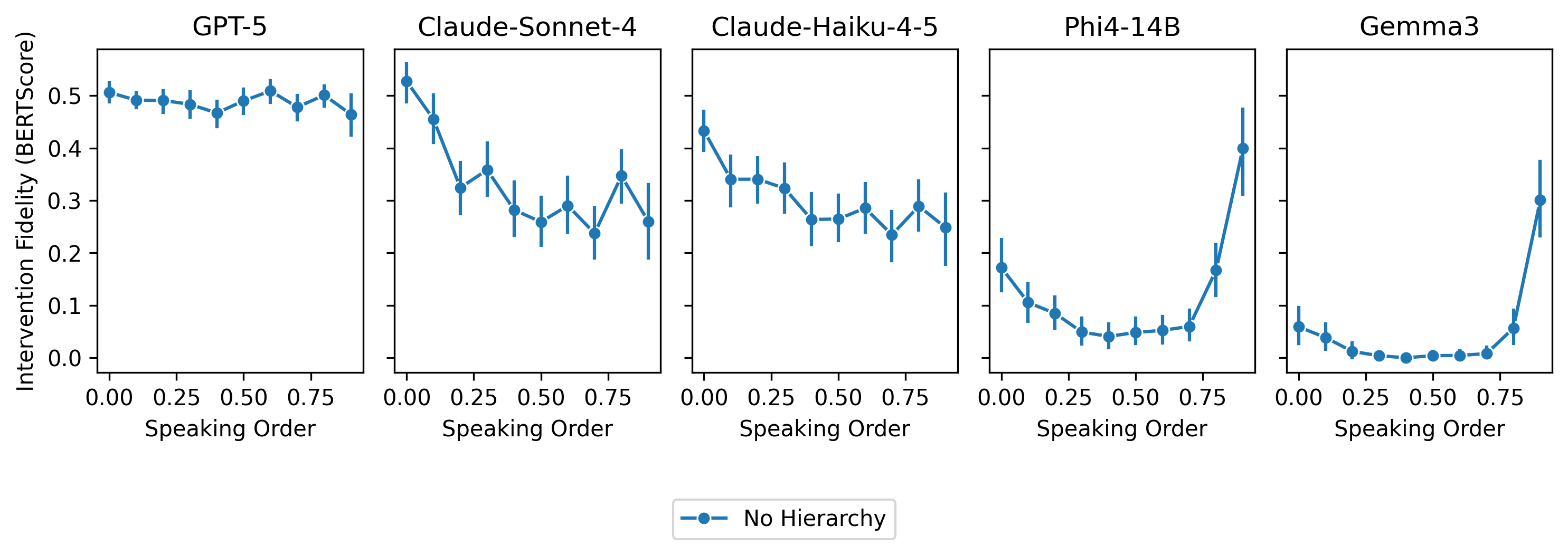}
\caption{\textbf{Speaker Order Bias.} \textit{BERTScore} measures the similarity between a speaker's reconstructed intervention summary $\hat{s}$ (based on the information in the debate summary), and their original intervention $i$. Relative Speaker Order ($\frac{k}{n}$) represents the temporal position of their intervention in the debate ($k$) adjusted for the number of interventions in the debate ($n$) with $\frac{k}{n}\approx0$ representing the earliest speakers in the debate and $\frac{k}{n}=1$ representing the last intervention.} \label{fig:lost_in_the_middle}
\end{figure}

{\textbf{Hypothesis.}} Existing research indicates that LLMs do not attend equally to all parts of the context when generating outputs (see Section~\ref{sec:background_bias}). In light of this, in Figure~\ref{fig:lost_in_the_middle} we assess the extent to which the temporal position of a speaker in the debate (that is, when they speak and thus where they occur in the LLM context) affects how much, and how accurately, the debate summary attends to that speaker and their contribution. We distinguish between a speaker’s \textit{temporal position} in the debate, defined as the order in which they speak, and their ideological position, referring to the stance they express on a given issue. To avoid confusion, we refer to the temporal position as \textit{speaking order}. We hypothesise that, in line with established findings on context utilisation in LLMs, speakers who contribute in the middle of a debate will be represented less accurately in the summary than those who speak earlier. Furthermore, we hypothesise that this speaking order bias will be mitigated by hierarchical summarisation methods, which decompose the task into a series of simpler extraction and aggregation steps.

\begin{figure}[!t]
\centering
\includegraphics[width=\textwidth]{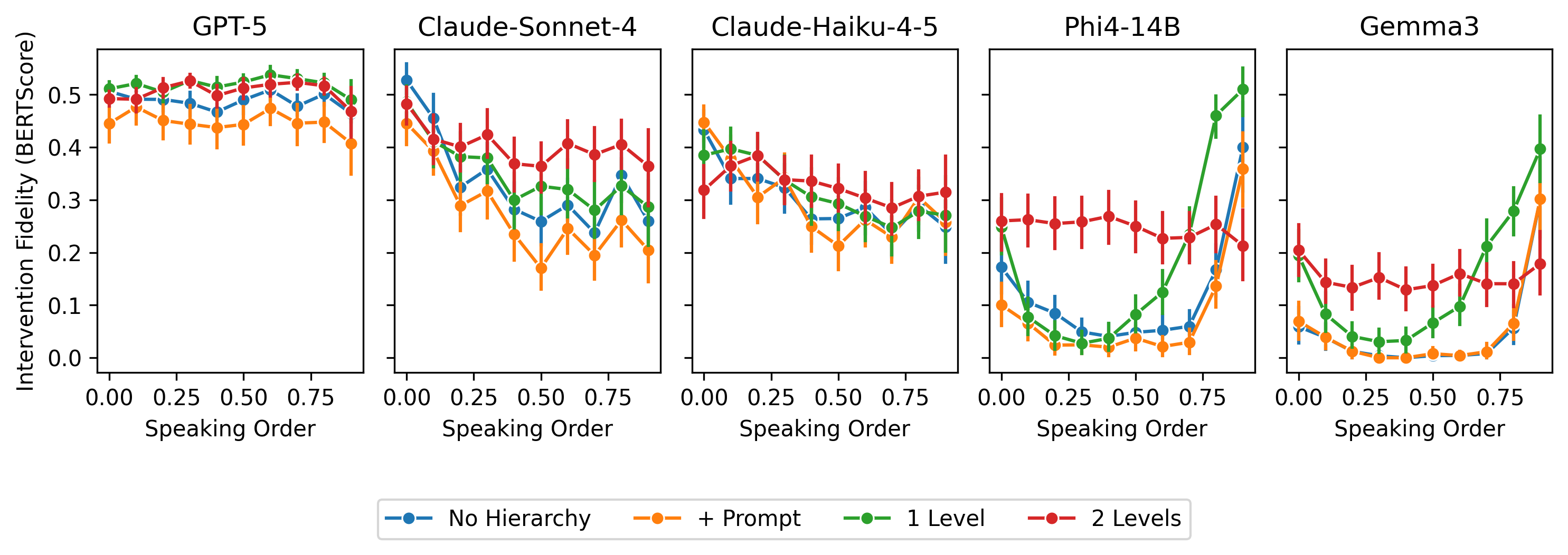}
\caption{\textbf{Speaking Order Bias -- Mitigating the lost-in-the-middle problem.} \textit{BERTScore} measures the similarity between a speaker's reconstructed intervention summary $\hat{s}$ (based on the information in the debate summary), and their original intervention. Relative Speaker Order represents the temporal position of their intervention in the debate, with lower scores representing earlier contributions.} \label{fig:speaker_order_bias_by_method_and_model}

\end{figure}


\begin{table}[!t]
\centering
\begin{tabular}{ll*{5}{S[table-format=-1.3]}}
\toprule
 & & {Sonnet-4} & {GPT-5} & {Haiku-4-5} & {Phi4-14B} & {Gemma3} \\
\midrule
Intercept & $\beta_0$ & -0.174* & -0.016 & -0.620** & -1.842** & -2.369** \\
Linear speaking order & $\beta_1$ & -4.410** & -0.841* & -2.821** & -2.761** & -1.830** \\
Quadratic speaking order & $\beta_2$ & 3.311** & 0.438 & 2.041** & 3.143** & 2.134** \\
+ Prompt (method effect) & $\gamma_P$ & -0.178 & -0.728** & 0.185 & -0.223* & 0.024 \\
1 Level (method effect) & $\gamma_{H_1}$ & -0.078 & -0.113 & -0.043 & 0.281* & 0.391** \\
2 Levels (method effect) & $\gamma_{H_2}$ & -0.319* & -0.237* & -0.423** & 0.084 & 0.187 \\
Order × + Prompt & $\delta_P$& 0.039 & 0.422 & -0.994 & 0.631 & -0.094 \\
Order × Hierarchical (1 Level) & $\delta_{H_1}$ & 0.989 & 1.321** & 0.608 & -2.380** & -1.794** \\
Order × Hierarchical (2 Levels) & $\delta_{H_1}$& 2.637** & 1.443** & 2.095** & 3.018** & 1.230* \\
Order² × + Prompt & $\theta_{P}$ & -0.055 & -0.422 & 0.947 & -0.610 & 0.096 \\
Order² × Hierarchical (1 Level) &$\theta_{H_1}$ & -0.950 & -0.991* & -0.620 & 3.744** & 2.532** \\
Order² × Hierarchical (2 Levels) &$\theta_{H_2}$ & -2.012** & -1.146* & -1.680** & -3.559** & -1.614** \\
\bottomrule
\end{tabular}

\caption{\textbf{Speaker order bias coefficients.} The beta regression coefficients for measuring speaker order bias and the mitigating effects of different generation methods.}
\label{tab:beta_regression}
\end{table}

\vspace{0.15em}\noindent{\textbf{Results.}} From Figure~\ref{fig:lost_in_the_middle}, it is clear that all models, with the exception of GPT-5, exhibit some speaker order bias. The Claude models attend more accurately to speakers who spoke earlier in the debate, whereas Gemma3 and Phi-4 favour those at the end. In the case of Gemma3 and Phi-4, we observe that many of the speakers from the middle of the debate are ignored entirely ($F1_{\text{BERT}}\approx 0$). 
In Figure~\ref{fig:speaker_order_bias_by_method_and_model} we plot speaker order bias for all summarisation methods and for each model. The figure indicates that the choice of generator method affects the extent of this bias.
To evaluate these effects we estimate a beta regression model. The dependent variable is intervention fidelity (\textit{BERTScore F1}), and predictors include relative speaker order $x_i$ (representing when the intervention occurred in the debate), its squared term $x_i^2$ (to capture non-linear patterns), indicators for each generator method (i.e., $\text{Method}_{m}$ is a dummy variable indicating if the debate was summarised using method $m$), and interactions to measure the effect of generator method on the speaker order bias. The regression is described by
\[
\begin{aligned}
\text{logit}(\mu_i) =\ 
& \beta_0 
+ \beta_1 x_i 
+ \beta_2 x_i^2 \\
& + \sum_{m \in \text{Methods}} \left[
    \gamma_{m} 
  + \delta_{m}  x_i
  + \theta_{m}  x_i^2
\right] \cdot \text{Method}_{m}
\end{aligned}
\label{eq:beta_regression_1}
\]
where $\beta_0$, $\beta_1$ and $\beta_2$ represent coefficients for the intercept, linear and quadratic terms, while $\gamma_m$ represents the method effect for method $m$, with $\beta_m$ and $\theta_m$ describing the effects of method $m$ on the speaker order bias. 
Table~\ref{tab:beta_regression} reports regression coefficients for all LLMs. All models display a strong positional bias under default flat summarisation. Most show a significant negative effect associated with speaking later in the debate ($\beta_1$). All models except GPT-5 also exhibit a significant positive quadratic term, $\beta_2$, indicating that summaries favour speakers who contribute early and late, to the detriment of those in the middle. As a naive baseline for mitigating attention bias, we evaluate a simple addition to the system prompt (\texttt{You must give equal attention to all speakers in your summary}). Across all models, this prompting strategy has no significant impact on speaker order bias.

Our proposed \textit{hierarchical} summarisation methods significantly alters the bias profile across all LLMs with greater effects at deeper hierarchical structures. The coefficient $\delta_H$ consistently shows a positive interaction with speaker order, reducing the bias towards earlier speakers, and in all cases where models show a bias towards later speakers (i.e., $\beta_2$ is significant and positive), we find a significant negative $\theta_H$ term. These results indicate hierarchical summarisation reduces speaker order bias and mitigates the so-called ``\textit{middle-curse}''.

\subsection{Lost in Translation}
\label{sec:results_language_bias}

\begin{figure}[t]
\centering
\includegraphics[width=0.7\textwidth]{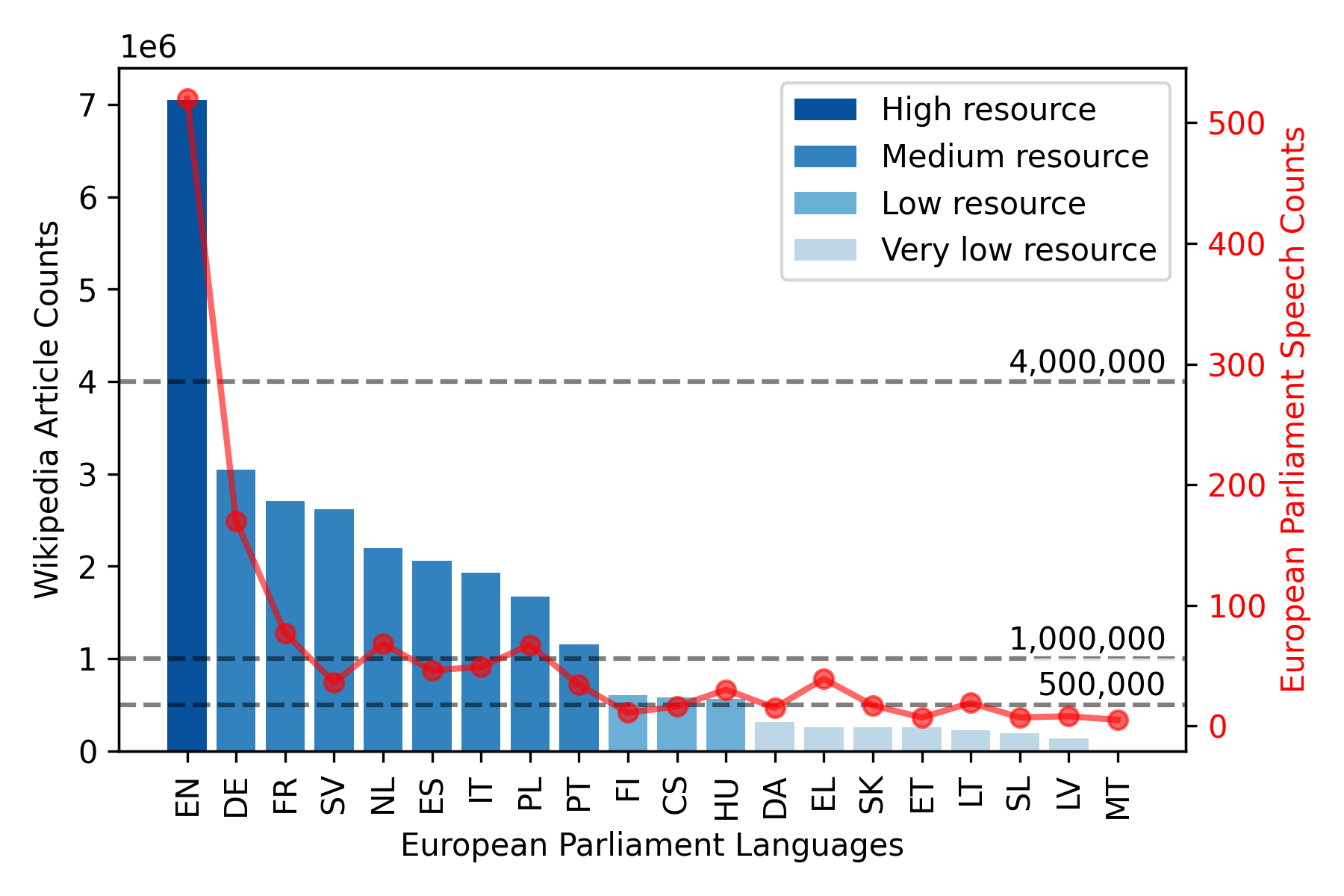}
\caption{\textbf{European Parliament Languages.} Our categorisation of language resource levels is based on the number of \textit{Wikipedia} articles available in each language. We find broad agreement between the prevalence of each language on \textit{Wikipedia} and its prevalence in our EP speech sample.} 
\label{fig:language_counts}
\end{figure}

\begin{figure}[t]
\centering
\includegraphics[width=\textwidth]{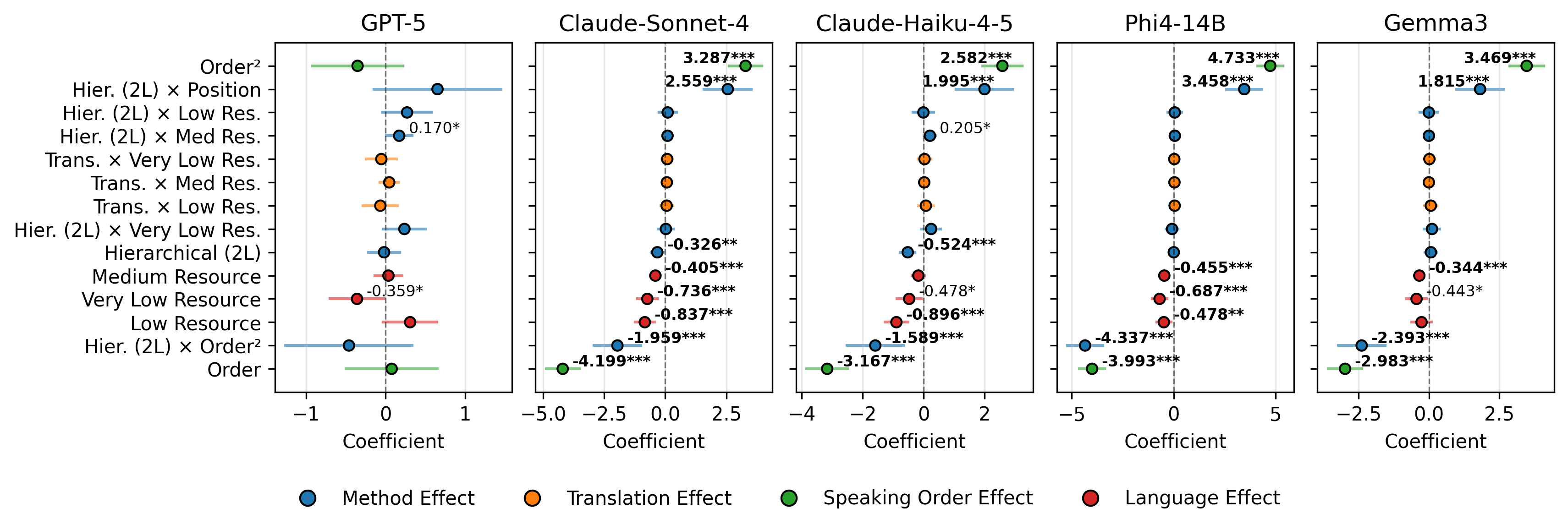}
\caption{\textbf{Modelling language effects alongside position effects.} Plots show the regression coefficients used to estimate intervention fidelity -- i.e., how accurately a speaker's intervention will be represented in the final summary. Coefficients include estimates of (i) position effects, indicating when the speech occurred in the debate, (ii) language effects, representing the resource-level of the language spoken, (iii) translation effects, specifying whether the intervention was translated into English prior to summarisation, and (iv) method effects, showing the effect of hierarchical summarisation on intervention fidelity, and it's interaction with the other terms.} \label{fig:cross_lingual}
\end{figure}

{\textbf{Hypothesis.}}  To investigate the effect of language spoken on intervention fidelity, we extend the regression from the previous section. First, we augment our data to include a many-to-one cross-lingual setting. Specifically, we generate two English summaries for each debate: one from the English transcript of the debate, and one from the original multilingual transcript. Second, we estimate how well each language was resourced in the training data of the LLMs.  As the specific training data for such models is proprietary, we use the number of articles on \textit{Wikipedia}\footnote{\url{https://www.wikipedia.org}} as a proxy. Figure \ref{fig:language_counts} shows the number of \textit{Wikipedia} articles available in each of the official EP languages in our dataset, and our heuristic categorisation of language resource level. Third, we include indicator variables in our regression to encode the language resource level (high, medium, low, very-low) and whether or not the summary was generated from the translated text. 
The results of this regression analysis are presented in Figure \ref{fig:cross_lingual}. As before, the beta regression estimates intervention fidelity, measured as \textit{BERTScore}, by comparing the original intervention $i_i$ with the information extracted from the debate summary pertaining to that speaker (the reconstructed intervention $\hat{s_i}$). In both cases (whether using the official EP translations or cross-lingual summarisation), we use the original intervention text, in its original language, as the source material for comparison ($i_i$), and we implement the \textit{BERTScore} metric using a multilingual embedding model, \texttt{sentence-transformers/paraphrase-multilingual-MiniLM-L12-v2}~\cite{reimers2019sbert}.

In the context of summarising plenary debates in the European Parliament, we hypothesise that interventions in medium, low, and very-low resource languages will be less faithfully represented in summaries than English interventions. Secondly, we expect that this effect will not be seen in the translated data, (i.e., where no cross-lingual summarisation is required). Finally, we hypothesise that hierarchical summarisation may mitigate this effect, as the summarisation process is broken down into simpler one-to-one cross-lingual summarisation steps prior to aggregation.

\vspace{0.15em}\noindent{\textbf{Results.}} Figure~\ref{fig:cross_lingual} shows that the speaker order effect is consistently stronger than any language effect and that, as before, speaker order bias is substantially reduced by hierarchical summarisation. In terms of language bias, we find that across all models there is a penalty associated with speaking in lower-resource languages and, contrary to our hypothesis, this effect arises regardless of the method of translation (i.e.\ whether using the official European Parliament translations or cross-lingual summarisation).
For their official translations, the European Parliament uses `pivot' or `relay' languages, where text is first translated into a third, more common language as an intermediate step \cite{amponsah2021rely}. The low scores associated with officially translated texts may result from inaccuracies or information loss in the original EP translations, which are propagated through the relay language and subsequent summarisation, and become apparent when compared with the source material in its original language. Across all our evaluations, we face challenges for low and very-low resource languages, due both to the scarcity of speeches in these languages in our dataset and to inherent limitations of multilingual embedding models.

\begin{figure}[t]
\centering
\includegraphics[width=0.75\textwidth]{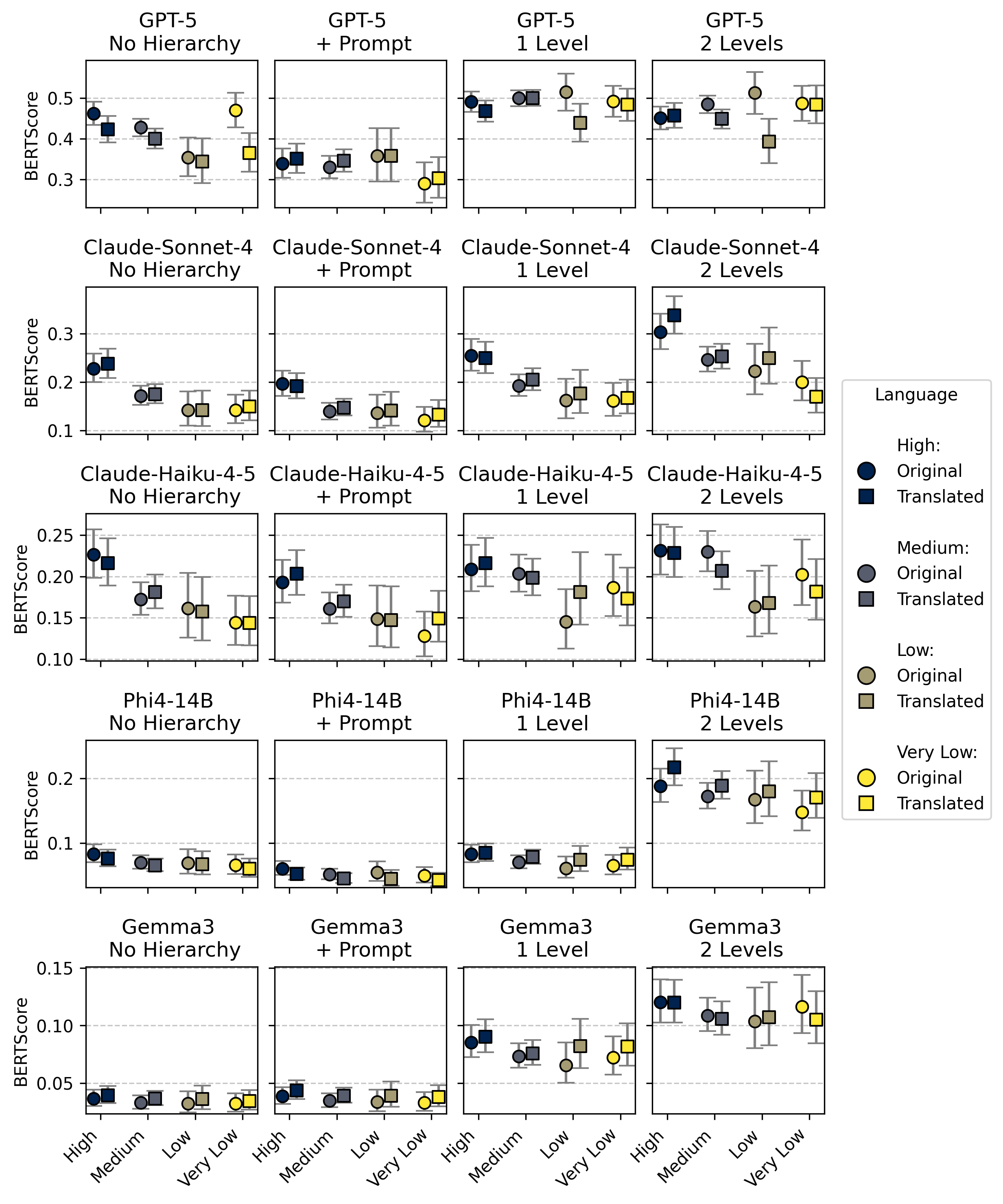}
\caption{\textbf{Marginal mean intervention fidelity by language resource level.} Plots show the mean predicted \textit{BERTScore} for an intervention made in the middle of a debate by speakers in different languages. Each debate is summarised into an English summary twice: once from the official English translation of the transcript, and once from the original multi-lingual transcript. Error bars indicate 95\% confidence interval. Higher scores indicate that the issues, positions, arguments, and proposals, made by members of that political group were more accurately/clearly communicated in the debate summary.} \label{fig:marginal_means_by_language}
\end{figure}

To better understand how language biases interact with different summarisation strategies, and to isolate these effects from speaker order bias, we plot marginal means for intervention fidelity assuming a speech in the middle of a debate in Figure~\ref{fig:marginal_means_by_language}. 
Here we observe that no language bias is detectable in the smaller models (\textit{Phi4} and \textit{Gemma3}), since intervention fidelity for speeches in the middle of the debate is already extremely low. In contrast, in the Claude models (\textit{Sonnet-4} and \textit{Haiku-4.5}), there is a clear penalty associated with speaking non-English languages, and, with the exception of medium-resource languages for \textit{Haiku-4.5}, there is no evident benefit from hierarchical summarisation.

\subsection{Lost in Partisanship}
\label{sec:results_party_bias}

{\textbf{Hypothesis.}} In addition to speaking order bias and language bias, LLM-based summaries may also differ in how they represent the issues, positions, arguments, and proposals put forward by speakers from different party groups. Given that all summaries exhibit some speaking order bias, we control for where interventions occur in the context when comparing intervention fidelity (\textit{BERTScore}). As in previous experiments, we fit a beta regression to model fidelity ($\textit{BERTScore}(i_i,\hat{s}_i)$) as a function of the speaker's turn $x_i$ (when they speak in the debate), its square $x_i^2$ (to capture the U-shaped attention pattern), and indicator variables for each of the party groups in the European Parliament (e.g.\ $\text{Party}_{A,i}$ is a binary variable indicating whether $\text{speaker}_i$ is a member of group $A$).
\begin{align*}
\text{logit}(\mu_i) =\ 
& \beta_0 
+ \beta_1 x_i 
+ \beta_2 x_i^2 
+ \beta_3 \text{Party}_{A,i} 
+ \beta_4 \text{Party}_{B,i}
+ \dots
\end{align*}
We centre the speaker order variable such that $x_i = 0$ refers to a speaker in the middle of a debate, and select the largest party (the European People's  (PPE-DE) in the case of the EP) as the reference group. 

In this case, we hypothesise that, given established evidence of partisan bias in language models~\cite{huang2024biasa}, speakers may experience different levels of intervention fidelity depending on their political affiliations.

\begin{figure}[t]
\centering
\includegraphics[width=0.75\textwidth]{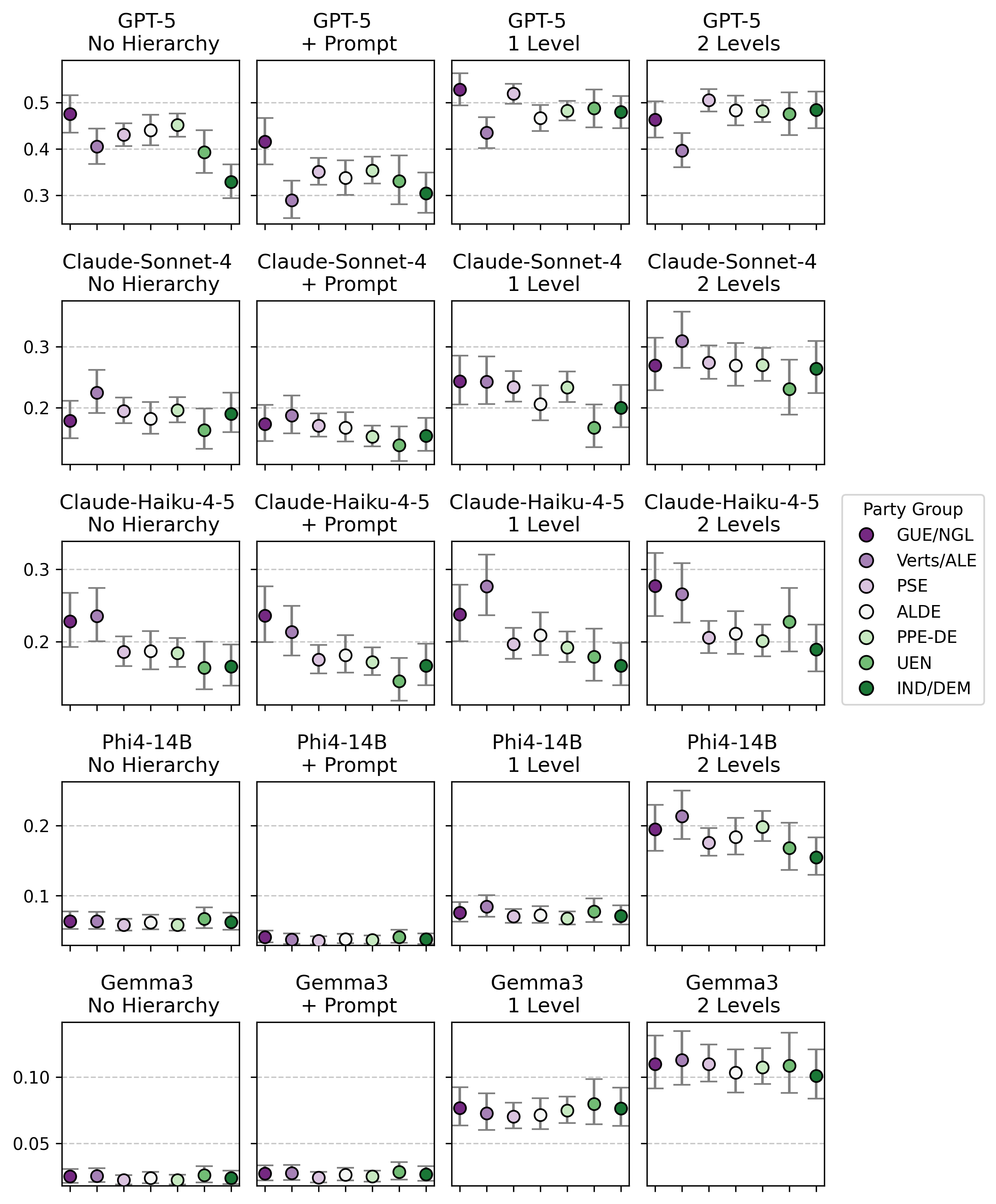}
\caption{\textbf{Marginal mean intervention fidelity by group.} Plots show the mean predicted \textit{BERTScore} for an intervention made in the middle of a debate by speakers of different EP party groups. Error bars indicate 95\% confidence interval. Higher scores indicate that the issues, positions, arguments, and proposals, made by members of that political group were more accurately/clearly communicated in the debate summary.} \label{fig:marginal_means_by_group}
\end{figure}

\vspace{0.15em}\noindent{\textbf{Results.}} Figure \ref{fig:marginal_means_by_group} plots the predicted marginal means for each EP party group across the different debate summarisation methods. The predicted marginal fidelity scores indicate that the choice of summarisation method can impact how well the summary reflects the positions and arguments of speakers from different political groups.
For the \textit{Gemma3} and \textit{Phi-4} models, the flat summarisation performed too poorly to identify any significant bias between different party groups, as too many speakers are ignored entirely based on their speaking order. However, in the case of the hierarchical methods -- which attend more equally to all regions of the debate -- we find that the Phi-4 summaries demonstrate a representation bias in favour of interventions made by left-of-centre parties. This partisanship was only measurable after we addressed the speaking order bias. 
A similar bias in favour of left-of-centre positions is apparent across all \textit{Claude} models. 

\begin{figure}[t]
\centering
\includegraphics[width=0.75\textwidth]{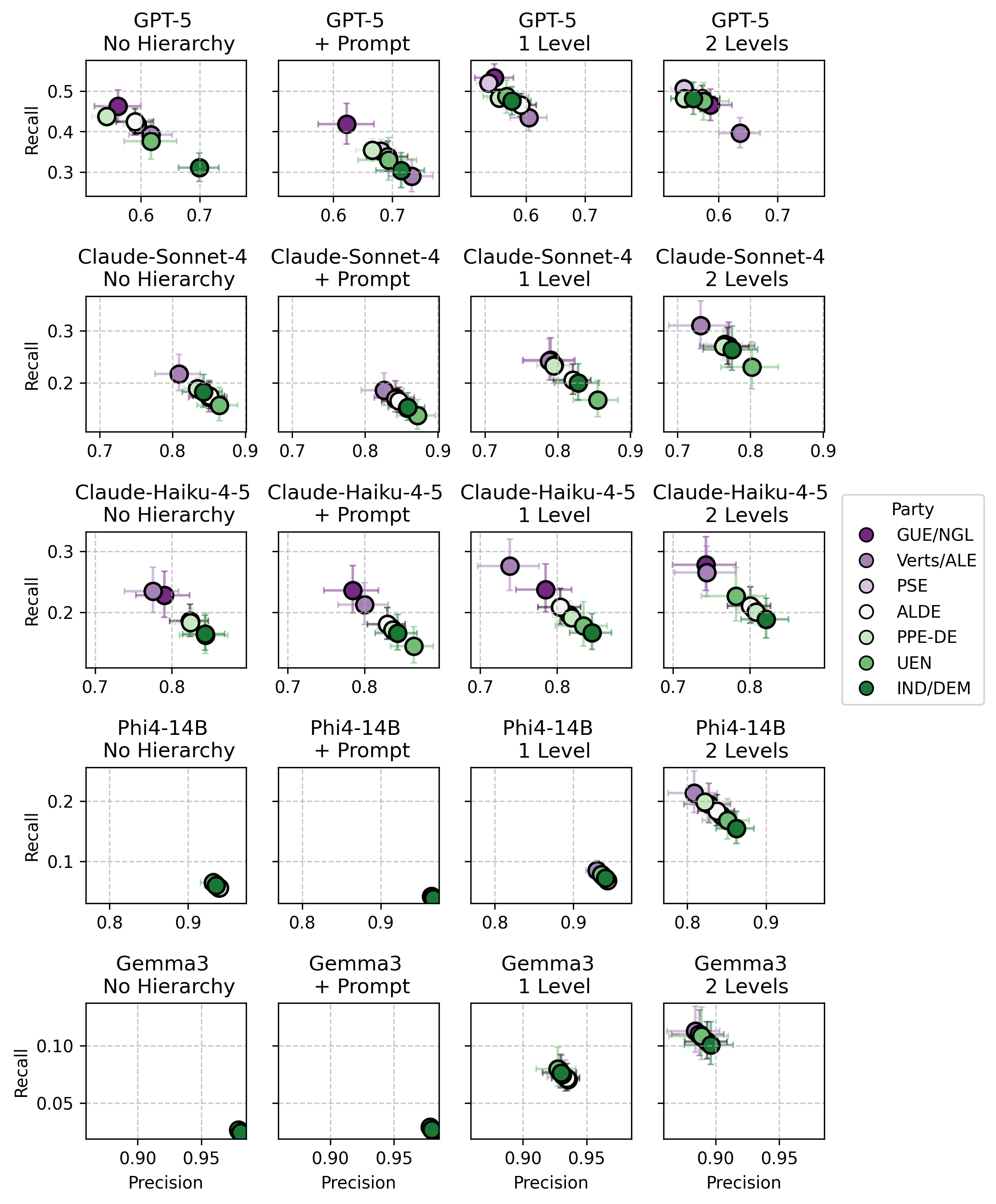}
\caption{\textbf{Marginal mean intervention precision and recall by group.} Plots show the mean predicted \textit{BERTScore} Precision, and \textit{BERTScore} Recall for an intervention made in the middle of a debate by speakers of different EP party groups. Error bars indicate 95\% confidence interval. Here the x-axis indicate the how \textit{precisely} the issues, positions, arguments, and proposals, made by members of that political group were communicated in the debate summary. In other words, \textit{how much of the information that is attributed to that speaker in the summary is supported by their original intervention?}.  Penalties on the x-axis would indicate a speaker's contributions were hallucinated or misrepresented. Conversely, the y-axis represents the \textit{recall} for the same contributions. In other words, \textit{How much of the speaker's original intervention is communicated (and clearly attributed to them) in the debate summary?}
As such, penalties on the y-axis indicate that a speaker's contributions were ignored.} \label{fig:precision_recall}
\end{figure}

We can explore this type of bias further by decomposing the \textit{BERTScores} into precision and recall components. Figure~\ref{fig:precision_recall} plots the marginal means for \textit{BERTScore} precision and \textit{BERTScore} recall for each party group. As such, significant gaps on the y-axis are indicative of inclusion bias, as less of the content in contributions made by the penalised can be found in the final summary. Similarly, gaps on the x-axis are indicative of hallucination bias, as the summary attributes contributions to the penalised groups that are less well supported by the source material. We must note here that, by default, implementations of \textit{BERTScore} give precision $=0$, and recall $=0$ in the case where the candidate material (in our case, the reconstructed summary $\hat{s}_i$) is empty. This is not consistent with typical definitions of precision, and as such, we award a precision of $1$ (and recall $=0$) to empty reconstructions (i.e., when no information relating to a given speaker could be extracted from the debate summary). Accordingly, Figure~\ref{fig:precision_recall} shows a strong precision/recall trade-off. Moreover, the left-of-centre parties that received better outcomes with respect to overall \textit{BERTScore} (especially from the \textit{Claude} models), are seen to have better recall scores, indicating that they benefit from an \textit{inclusion bias}, with more information from their contributions making it into the final summaries, when compared to other parties.

\subsection{Key Findings}
The experimental results presented in Sections \ref{sec:results_order_bias}--\ref{sec:results_party_bias}, spanning five LLMs and multiple summarisation methods, reveal three distinct forms of bias in parliamentary debate summaries:
\begin{enumerate}
    \item Positional bias affects all models under flat summarisation, with speakers in the middle of debates systematically under-represented. Hierarchical summarisation consistently addresses this problem, flattening the U-shaped attention pattern across all tested models.
    \item Language-based representation disparities are apparent across all models and persist regardless of method: speakers in lower-resource languages receive less faithful representation even when hierarchical methods are applied, and this penalty appears whether summaries are generated from multilingual transcripts or from official EP translations. Notably, when modelled alongside the speaking-order effects, these speaking-order effects dominate between-language effects. 
    \item Most models demonstrate a partisan bias, which manifests through inclusion/exclusion. Results indicate that interventions made by speakers from left-of-centre parties are more completely represented in the summaries than others. In the smaller open-weight models, this bias was not measurable without implementing hierarchical summarisation, as the baseline exclusion rate was too high to measure any between-party effects.   
\end{enumerate}

\section{Conclusions and Future Work}

Accessible summaries of parliamentary debates can enhance transparency, thereby strengthening the connection between the public and the democratic institutions that represent them. LLMs offer a potentially promising approach for this task, given their demonstrated ability to generate well-structured and coherent summaries across a range of domains. However, these models also raise concerns, as they can exhibit both algorithmic \cite{liu2024lost,ravaut2024context} and social biases \cite{bartl2024showgirls}, which must be evaluated in parallel with more conventional metrics of summary quality and accuracy. While many established methods exist for evaluating automatically-generated summaries, they are often insufficient for assessing attribution accuracy in political debate contexts. In particular, it is crucial not only that a summary faithfully reflects the arguments, positions, and proposals expressed in the debate, but also that these elements are correctly attributed to the speakers who presented them. 

To address this gap, we have introduced a structured framework for generating and evaluating political debate summaries. Our framework focuses on the substantive content of each intervention, namely issues, positions, arguments, and proposals, and evaluates how accurately these components are attributed to speakers and communicated in the final summaries.
By applying this framework in the context of plenary speeches from the European Parliament, we have identified three forms of bias present in LLM generated summaries. 
We find that speakers can be systematically excluded from debate summaries on the basis of (i) \textit{when} they speak, (ii) \textit{what language} they speak, and (iii) their \textit{political affiliations}:

First, speaker-order bias affects all tested models under flat summarisation: speakers in the middle of debates are systematically under-represented, consistent with established findings on context utilisation in LLMs. We address this through hierarchical summarisation methods that impose structure based on political science research on policy bargaining and debate. Unlike previous hierarchical approaches that failed when applied to unstructured documents, our method prescribes a scaffolding derived from the substantive elements of debate -- issues, positions, arguments, and proposals. This structured approach effectively mitigates the lost-in-the-middle problem across all tested models.

Second, we observe persistent language-based representation disparities. Speakers using lower-resource languages receive less faithful representation regardless of summarisation method, and this penalty emerges even when summaries are generated from official European Parliament translations rather than through direct cross-lingual summarisation.

Third, the reduction in speaking order bias achieved through hierarchical methods reveals differences in partisan representation that were previously obscured by severe attention effects. Particularly in Claude models, we observe that left-of-centre party interventions show higher recall, indicating more complete coverage in summaries compared to other political groups.

Taken together, these results demonstrate that while architectural modifications to summarisation workflows can address attention-related biases, they cannot compensate for biases rooted in training data or language resource disparities. Integrating these insights into AI-driven platforms for parliamentary transparency will require careful consideration of how different biases interact and compound. This work underscores the need for nuanced, domain-specific evaluation of LLM applications.

\backmatter

\bmhead{Supplementary information}

If your article has accompanying supplementary file/s please state so here. 

Authors reporting data from electrophoretic gels and blots should supply the full unprocessed scans for key as part of their Supplementary information. This may be requested by the editorial team/s if it is missing.

Please refer to Journal-level guidance for any specific requirements.

\bmhead{Acknowledgements}

This work was partially supported by Research Ireland grant number SFI/12/RC/2289\_P2.  

\section*{Declarations}

\bmhead{Competing Interests} 
The authors declare no competing interests.

\bmhead{Funding}
This work was partially supported by Research Ireland grant number SFI/12/RC/2289\_P2.  

\bmhead{Data Availability}
All data are available at \url{https://github.com/parliview/debate-summ-cross-lingual}.

\bmhead{Code \& Materials} 
All code and materials for (i) experiments are available at \url{https://github.com/parliview/debate-summ-cross-lingual}, and (ii) annotation/validation task are available at \url{https://github.com/eoghancunn/position_extraction_annotation}.

\bmhead{Ethics Approval} 
N/A.

\bmhead{Author Contribution} 
EC, JC, and DG designed the study; EC conducted the experiments, EC, JC, and DG analysed the results, and all authors participated in writing the manuscript.



\bibliography{references}

\end{document}